\documentclass[12pt]{article}

\usepackage[pdftex]{graphicx}
\DeclareGraphicsExtensions{.pdf,.jpeg,.png}

\begin{document}

\title{Wrapped Classifier with Dummy Teacher for\\ training physics-based classifier at\\ unlabeled radar data}

\author{Oleg~I.Berngardt, Oleg~A.Kusonsky,\\ Alexey~I.Poddelsky, Alexey~V.Oinats}

\maketitle

\begin{abstract}
In the paper a method for automatic classification of signals received by EKB and 
MAGW ISTP SB RAS coherent scatter radars (8-20MHz operating frequency) during 2021 
is described. The method is suitable for automatic physical interpretation of the 
resulting classification of the experimental data in realtime. We called this 
algorithm Wrapped Classifier with Dummy Teacher. The method is trained on unlabeled 
dataset and is based on training optimal physics-based classification using clusterization results. 
The approach is close to optimal embedding search, where the embedding is interpreted as a vector of 
probabilities for soft classification. The approach allows to find optimal classification algorithm, 
based on physically interpretable parameters of the received data, both obtained during physics-based 
numerical simulation and measured experimentally. Dummy Teacher clusterer used for labeling unlabeled 
dataset is gaussian mixture clustering algorithm. 

For algorithm functioning we extended the parameters obtained by the radar with additional parameters, 
calculated during simulation of radiowave propagation using ray-tracing and IRI-2012 and IGRF models 
for ionosphere and Earth's magnetic field correspondingly. 

For clustering by Dummy Teacher we use the whole dataset of available parameters (measured and simulated ones). 
For classification by Wrapped Classifier we use only well physically interpreted parameters. As a result we 
trained the classification network and found 11 well-interpretable classes from physical point of view in 
the available data. Five other found classes are not interpretable from physical point of view, demonstrating 
the importance of taking into account radiowave propagation for correct classification.
\end{abstract}

\section{Introduction}\label{sec:introduction}

The problem of classifying multidimensional data is quite complex,
intensively studied and finds its applications in various fields, including
geophysics \cite{Yu_2021}. One of the informative diagnostic tools for the
magnetosphere, ionosphere and upper atmosphere are SuperDARN (Super
Dual Auroral Radar Network) and similar decameter radars. Currently there are more than
35 such an instruments around the world \cite{Nishitani_2019}. 
The radar transmit sounding pulses and receive scattered signal used to study the 
scattering irregularities.
The high data amount produced
by these radars, however, is accompanied by the difficulty in interpreting
these data. The received radar signals are a mixture of signals formed
by various physical mechanisms that are combinations of radiowave reflection
and scattering processes - scattering from the ground and sea surfaces,
refraction and focusing in the ionosphere, scattering
by ionospheric irregularities of E and F layers, aligned
with the Earth's magnetic field, scattering 
on meteor trails, mesospheric radio echo, etc \cite{Nishitani_2019}. For the
statistical and morphological analysis of such phenomena, one needs
to automatically identify and separate
these signals. In this problem, different methods are
used, based on the analysis of statistically processed data\cite{Blanchard_2009},
on machine learning methods\cite{Ribeiro_2011}, and on the analysis of individual
soundings before their statistical processing\cite{Lavygin_2019}.

Since 2012, ISTP SB RAS operates the EKB coherent
scatter radar in the Sverdlovsk region, and since 2020 it operates the
MAGW radar in the Magadan region. The radars are similar
to SuperDARN radars. Unlike the mid-latitude EKB radar, which
observe only several types of scattered signals, mostly groundscatter \cite{Berngardt_2015},
the higher latitude MAGW radar observes more signal types, which are difficult
to identify, interpret and analyze. This cause the problem of automatic identification of
scattering signal types at ISTP SB RAS radars. 

Frequently, such a task is solved manually, using the analysis of
specific cases\cite{Blanchard_2009}, or by statistical methods or machine learning
methods, formulating separation rules based on datasets of labeled
examples and/or using initial assumptions 
about number of classes and separation surface shape\cite{Ribeiro_2011,Lavygin_2019}. 
In this paper, we propose an identification method
based on modern machine learning methods with using an unlabeled
dataset, and with a significant involvement of knowledge about possible
physical scattering mechanisms.

Usually, when separating data, one of two
approaches is used - clustering or classification. They
allows separating a set of points in a multidimensional space into
clusters/classes - subsets separated according some rules.

Clustering uses unlabeled dataset(we do not initially know actual clusters each point belongs to),
and some clustering method separates the data into a clusters. 
Clustering of new data requires either new training on an expanded
dataset, or uses an approaches to classify new the point based on classes
of already classified points from the training dataset (for example,
by K-Nearest Neighbors algorithm). There are many clustering methods exists, and changing
the method most often results to different separation of the
data into clusters. The advantage of this method is that there is no
need to label each point of the training dataset by its class.
In terms of machine learning, it does not require a teacher. The disadvantage
is the uncertainty in choosing the correct method for clustering new
points, as well as high hardware requirements when training on large
datasets and when used in real time.

Classification is usually carried out on the basis of training a certain
decision scheme (decision trees, neural networks, etc.) on a labeled
training dataset, and then applying the trained scheme to classify
new points from a new dataset. When classifying data, the important
problem is the presence of a labeled dataset, in terms of machine
learning - the presence of a teacher. The presence of a teacher allows
one to know the number of expected clusters (classes). Modern classification
methods allow either to predict the class of each point ("hard classification"),
or to predict the probabilities that this point belongs to one or
another class ("soft classification"). The predicted class of a point
is usually defined as the class with the highest probability. Thus,
the advantage of the classification are: known number of clusters,
easiness of interpretation of each class, fast and optimal algorithm for classifying
new points. The disadvantage is the need for a labeled training dataset.

The application of machine learning to the problem of data classification
of SuperDARN and similar radars is not new, but previously it was
limited to the simpler problem of separating signals scattered from
the earth's surface and signals scattered by the ionosphere \cite{Blanchard_2009}
using a labeled dataset\cite{Lavygin_2019} or clusterization\cite{Ribeiro_2011}. 

Our method combines classification and clusterization into a single scheme 
to train automatic classification of the data from the ISTP SB RAS coherent scatter
radars and provides interpretability of the results. The main idea of the method is using clustering for
all available data: experimental one and physics-based numerically simulated one, and training optimal
classifier network, that uses smaller number of dataset features to produce 
optimal classification to "hidden classes".
These hidden classes can be used as an input for another network that produce 
optimal prediction of clusterization results. The use of interpretable features 
for classification allows us to interpret the hidden classes 
from a physical point of view. This approach is close to "optimal
embedding" approach, widely used in natural language processing.

Since the two schemes - clustering and classification in this method
work together, we called such a classification scheme and its training
a "Wrapped Classifier with Dummy Teacher" , where transformation
of hidden classes into clusters acts as an auxiliary "Wrap"
that performs training of the "Classifier". 
Based on its function the clustering algorithm is called by us a "Dummy
Teacher".
The Wrap and Dummy Teacher
are not directly involved in processing new data,
but allows us to train the optimal classifier at train dataset. 

\section{Idea, network architecture and training}

To combine the approaches, we create a three-stage scheme that does
not require a teacher. 
At the first stage we calculate additional parameters from physics-based numerical 
simulation of radiowave propagation.
At the second stage, we create a labeled 
dataset from the original unlabeled dataset using Dummy Teacher 
clustering algorithm.
At the third stage, we create a classification scheme consisting
of two sequential neural networks - one of which (Classifier)
creates an optimal representation of the data 
as 20-dimensional vector (embedding) interpreted by us later as hidden classes
probabilities.
Second network (Wrap) transforms the
hidden classes probabilities into cluster numbers of the labeled dataset.
During training, the Classifier and Wrap will be jointly trained 
for best fit of labeled dataset.

We cannot initially justify which clustering scheme to use in the
first stage, and be sure of the adequacy of the clusters it creates. In this sense,
clustering acts as a dummy teacher who somehow clusters our data,
but make mistakes, clusters data incorrectly, selects more or less clusters
than necessary, does not merge clusters, uses non-physical variables for
clustering, and so on.

It conducts clustering from experiment to experiment (from day to
day, from radar to radar) independently: the same scattering types
in different experiments can have different cluster numbers. Therefore,
when interpreting clustering results 
we do not only separating and merging the clusters, but also
renumber them. We can not fully trust such a teacher,
and be sure of the correctness of his clustering. We only assume that,
in some sense, this clustering is approximately correct. That is why we call
such a teacher a Dummy Teacher. Based on the results of its clustering,
we build and train an optimal classifier that produce
optimal physics-based classification and has generalizing and interpreting capabilities
that are lacking in a Dummy Teacher.

One of the intensively developing directions of machine learning today
is natural language processing, having a huge amount of applications
- from automatic translation and text abstraction problems to chat
bots and question-answer systems\cite{Jurafsky_2009}.

One of the central places in natural language processing is the vector
representation of objects (embedding). Embedding represents
an object (word or N-gram) as a vector, usually of a fixed dimension.
A significant advance in the processing of natural languages is taken
by adaptive embedding, which was apparently presented for the first
time in word2vec procedure. It was built on the basis
of the requirement for an optimal vector representation of a word
to solve the problem of predicting a word by its neighbors\cite{Mikolov_2013}.
The optimal embedding constructed in this way has two useful properties:
the closer two words in their meaning, the closer Euclidean distance
between their embeddings; and replacing the words with
their found embeddings makes it possible to effectively 
predict a word by its neighbors. Thus, the optimal embedding reflects
the semantic load of words and widely used for text analysis.

More general application of the embedding is to represent the
meaning of sentences used in automatic translation algorithms (for
example, in seq2seq\cite{seq2seq}). In this case, two networks are trained, one of
which (Encoder) transfers to the other (Decoder) the meaning of the
sentence, represented in the form of a multidimensional vector, and
the algorithm for the calculation of this vector arises as a result
of solving the task of optimal translation of this sentence from 
one language to another.
In this case, sentences that are close in meaning turn out to be also
close in the Euclidean distance between their vectors
\cite{seq2seq}. Such a representation makes it possible to effectively
solve the problem of automatic translation of texts. Thus, the obtained
multidimensional vector can be interpreted as an embedding for sentences,
optimal for solving the problem of automatic sentence translation.

The problem of soft classification by a neural network also can be
considered as constructing an effective embedding for points: the
dimension of the embedding (output) vector of classes probabilities is equal to the number of
classes, and the points belonging to the same class should be close in terms
of the distance between their embedding vectors. This optimal representation
is at the same time a solution of the classification problem.

Therefore, the problem of automatic data classification in our case
can be reformulated as the problem of constructing of an effective
embedding for each data point (Classifier), 
which will allow us to
build a network (Wrap) that conducts classification that is optimally
close to the clustering produced by the Dummy Teacher. Such efficient
embedding can be interpreted as classifying the data into optimal (hidden)
classes. By choosing physically clear and easily interpretable inputs
for Classifier, these classes can also be interpreted from a physical
point of view.

\begin{figure}
\includegraphics[scale=0.22]{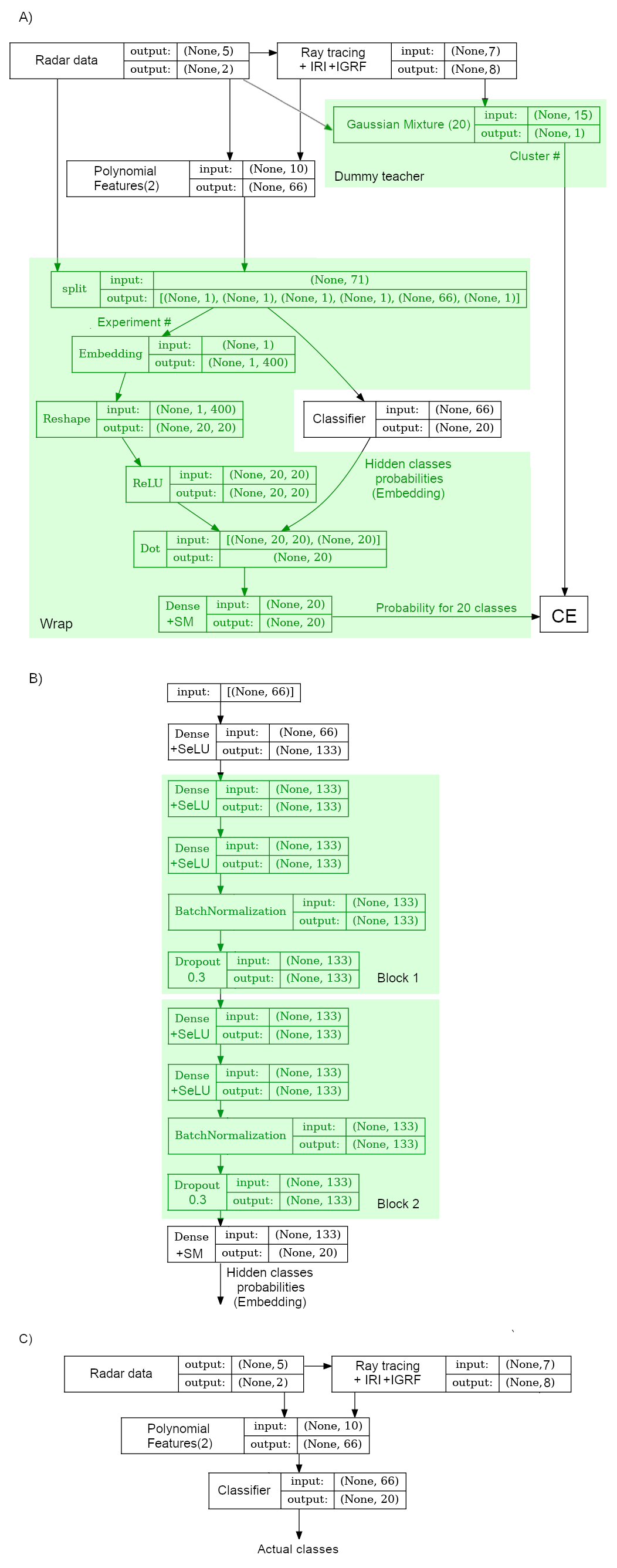}

\caption{Architecture of the Wrapped Classifier with Dummy Teacher and dimensions of vectors. 
A) - network architecture
in training mode; B) Classifier architecture C) - network architecture
in processing mode. "None" labels correspond to the number of records
in the dataset.}

\label{fig:1}
\end{figure}

We use the following architecture ("Wrapped Classifier"),
shown in Fig.\ref{fig:1}, consisting of two networks (Classifier
and Wrap), trained cooperatively and a clusterer (Dummy Teacher).
Once trained, we later use only the trained Classifier for classifying
new data.

From Fig.\ref{fig:1}A,C one can see that the network architecture
is ideologically close to the word2vec network architecture, but instead
of the embedding matrix for words, a neural network - Classifier is
used (Fig.\ref{fig:1}B), and the problem of predicting word by its context
is replaced by the problem of predicting Dummy Teacher cluster number by 
hidden classes probabilities. Classifier generates optimal embeddings and
takes into account the physics of the problem by using only physically
well-interpreted parameters for classification. This representation is
further converted by the Wrap into classification closest to clusterization
by Dummy Teacher. An important detail is that
the numbering of the clusters differs from experiment to experiment,
and the the clusters must be renumbered for each experiment. To do
this, the Wrap uses adaptive embedding, which, unlike traditional
vector-valued embeddings, is a non-negative matrix-valued embedding (implemented by sequential
use of the Embedding, Reshape and ReLU layers of TensorFlow library).
The experiment number is used as an index for the embedding (in our
case, this is a unique number constructed from the radar number and
the day number). The renumbering of classes is performed by matrix
multiplication of the classes obtained from the Classifier by a matrix
embedding that is unique for each experiment. Conceptually, this operation
is close to a very simplified Attention \cite{Vaswani_2017} mechanism. The
renumbered class obtained in this way goes to the decision layer.
This method allows us to speed up the learning process 
and the memory amount required for the network training.

We use a polynomial straightening space (Polynomial Features,
appends squares of parameters and all cross products of parameters)
at the input of the Wrapped Classifier. This caused by to the fact that 
weighted sum of the squares of the Doppler shift and the
signal spectral width is already used as a good criterion for the identification of groundscatter
signals \cite{Blanchard_2009}. So taking into account the squares of the input features 
and the cross products of the input features, as will
be shown below, increases the efficiency of the classification.

The Classifier is a fully connected network
consisting of input and output layers, and hidden layers combined
into two identical blocks: 2 fully connected layers + Normalization
+ 30\% Dropout (shown in Fig.\ref{fig:1}B). 
All the activation functions, except for the output
layer, are SeLU.

SeLU is used to reduce the effect of gradient fading, DropOut and
Normalization - to reduce overtraining of the network and to speed up
its training, respectively.

The number of neurons in each layer of the Classifier
per one exceeds doubled dimension of the input data and is chosen
in accordance with the Kolmogorov-Arnold theorem \cite{Arnold_1963}
for the most optimal representation of the input data.
Classifier has about 83000 trained parameters, 
and Wrap has about 206000 trained parameters. 

For training, the initial dataset (data from two radars for January-September
2021, \textasciitilde{} 3 million records) is splitted into training,
validation and test datasets in proportion of 64\%:16\%:20\%. The
weighted cross-entropy is used as a loss function, the stop condition is the
early stoping for 20 epochs based on the value of the AURPC metric
on the validation dataset.

The difference from traditional schemes is that the embedding of the
Classifier scheme is normalized - it is non-negative, and the sum
of its components is equal to 1, which is implemented by Softmax output
activation function. This allows interpreting the Classifier output
as the probability of belonging the point to a particular (hidden) class.

The proposed algorithm consists of 3 stages:

1) adding to the experimental data obtained by the radar, the physical
parameters calculated from these experimental data and physical models, 
making unlabeled dataset;

2) clustering each experiment of the resulting dataset by the Dummy
Teacher to obtain a labeled training dataset for the Wrapped Classifier.

3) joint training of the Wrapped Classifier network (Wrap and Classifier)
on labeled dataset to obtain an optimally trained Classifier from the condition of the
maximum coincidence between the classification made by Wrapped Classifier
and clustering made by the Dummy Teacher.

To speed up the network training each stage was carried out sequentially,
with saving the resulting datasets to storage. To interpret the new data,
only stage 1 is required, followed by the processing of the data received
from 1st stage by the trained Classifier. Let us describe the stages in details.

\subsection{Stage 1: adding physical parameters}

Dummy Teacher separates the data according to all the dataset features
available, not paying a special attention to the physicality
of these parameters - their suitability for interpreting the physical
mechanisms of formation of a particular type of scattered signal.
Such a processing that is not based on physical mechanisms will be
nearly useless later for the interpretation of the data. Therefore,
the Classifier does its classification based only on well-understandable
physical parameters included into the dataset, both measured and numerically simulated by physical models 
based on the parameters measured by the radar. This allows
us later to interpret the classes obtained by the Classifier from a physical
point of view.

The main parameters affecting the interpretation of the scattered
signals in the ionosphere are the position of the scattering
region on the radiowave propagation trajectory, the angle between
the propagation trajectory and the Earth's magnetic field at the scattering
point, and the scattering height\cite{Nishitani_2019}. It is also important
for consideration to take into account the hop-like nature of propagation
to understand at which hop of the trajectory the scattering occurred,
and how many times the radio signal was reflected from the ionosphere
and from the Earth's surface.

Therefore, based on the measured parameters of the received signal
and known experimental setup, we can 
calculate additional physical parameters using radiophysical
and geophysical models describing radiowave propagation in an inhomogeneous ionosphere.

Geometric optics (ray tracing) \cite{Ginzburg} was used to calculate the wave propagation
trajectory in the approximation of non-magnetized
ionosphere. Ionospheric refraction is calculated from the international reference
model of the ionosphere IRI-2012\cite{Bilitza_2014}. The Earth's magnetic field
is calculated from the international reference magnetic field model
IGRF \cite{IGRF_2015}. As input parameters for calculating the trajectory, we
used: the received signal elevation angle (it was assumed that it coincides
with the emission elevation angle) calibrated by the
method\cite{Berngardt_2020}; the azimuth of the radar beam; the operating frequency
of the radar; the geographic position of the radar;
date/time and radar range.

As the physical parameters characterizing the calculated propagation trajectory,
the following parameters were chosen, obtained by raytracing:

- the sine of the angle between the wave vector of the radiowave and
the horizon at the scattering point. It allows one to estimate how close
the scattering point to the point of the reflection from the ionosphere, where this parameter becomes zero.
Before the point of ionospheric reflection this parameter is
greater than zero, and after the point of reflection (this parameter
is less than zero. This allows to later interpret classes in terms
0.5 and 1.5 hop positions;

- the cosine of the angle between the wave vector and
the Earth's magnetic field at the scattering point. This parameter
allows one to determine whether the scattering is near the perpendicular
to the magnetic field (aspect sensitive scattering) or not. In case of scattering from magnetically
oriented inhomogeneties the parameter is close to zero;

- the sine of the angle between the wave vector and the
horizon in the middle of the path length to the scattering point. It
allows one to understand whether the signal is scattering
from the Earth's surface, because in this case this parameter should
be close to zero;

- the sine of the angle between the wave vector and the
horizon at a quarter of the path length to the scattering point.
As in previous case, it allows us to estimate whether the signal 
is the second scattering from the ground after two-hop propagation or not.

- the sine of the angle between the wave vector and the
horizon at three quarters of path length to the scattering point. It has
similar interpretation as the previous case.

- the hop number for hop propagation (the number of reflections from
the Earth's surface + 1). It allows one to additionally guess at which hop of
the trajectory the scattering occurs.

- the altitude at which the scattering occurs. It allows one to guess
about a possible mechanism of the scattering: the region of 70-100
km corresponds to meteor trail scattering, 100-130 km - scattering in the E-layer,
above 150 km - scattering in the F-layer, if the altitude close to
zero - most likely this is scattering from the Earth's surface. 
The heights greater than F2 maximum height (about 400km) usually should not occure,
so in this case propagation trajectory calculated incorrectly.

- the effective scattering height - the height obtained in the assumption
of straight-line (refraction-free) wave propagation. This parameter can help 
to interpret the
cases where refraction in the ionosphere can be neglected. For example,
the case of meteor trail scattering and scattering in the E-layer at
0.5 hop. 
In most often cases the interpretation of this parameter 
depends on sounding frequency, 
so using of this parameter could cause questions, and will be discussed
later.

In addition, the physical parameters characterizing the scattering
mechanism are the Doppler shift of the radiowave (associated
with the velocity of the scattering irregularities and propagation
in non-stationary ionosphere) and the spectral width of the signal
(associated with the lifetime of the scattering irregularities).

\begin{figure}
\includegraphics[scale=1.3]{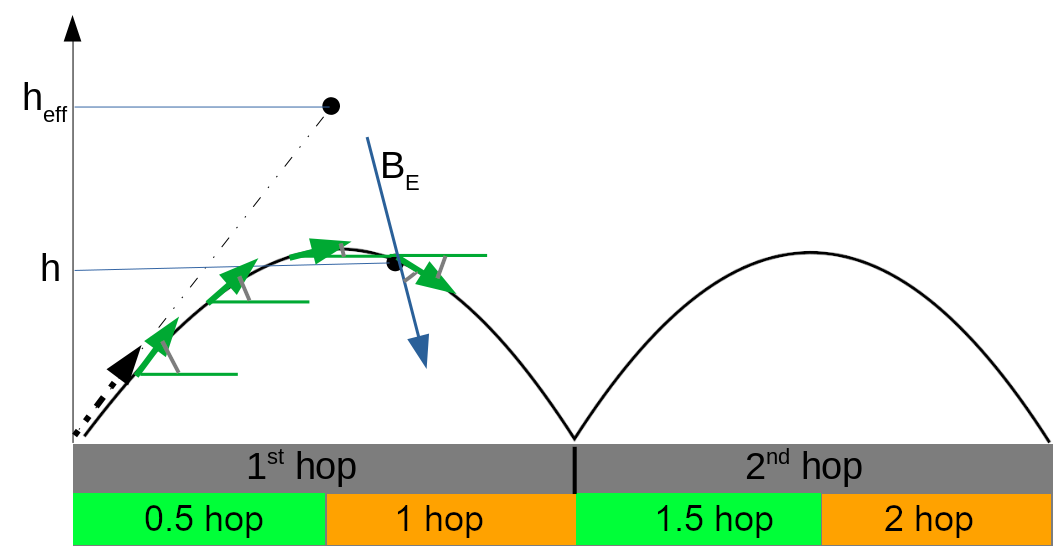}

\caption{Scheme for calculating physical parameters determined from numerical
simulations. The black solid line is the trajectory of the radio signal
propagation, calculated using ray-tracing. blue arrow - the direction
of Earth's magnetic field, green horizontal lines marks the direction
of the horizon, green solid arrows - the direction of propagation
of the radio signal. The half hops marked at the bottom (0.5,1,1.5,2)
are not used as parameters for the network, but are used later when
interpreting the classes.}
\label{fig:2}
\end{figure}

Thus, the classification model (Classifier) depends on 10 input parameters,
only two of which (Doppler shift and spectral width) were directly
measured by the radar, and the rest 8 of them were obtained using numerical
simulation based on the measured parameters of received signal. 
These parameters are illustrated in Fig.\ref{fig:2}.

The propagation trajectory of the radiowave is calculated by the
geometric optics method \cite{Ginzburg} based on the ionosphere model (given by
the IRI model) and the characteristics of the radio wave (the frequency
of the sounding signal and the angles of its arrival - azimuth and
elevation). The necessary smoothness of the ionosphere in calculations is provided
by fitting it with local B-splines of the 2nd order (parabolic interpolation
of electron density between spatial and altitude points given by the
IRI model). The radiowave is thought propagating in a plane determined
by the sounding azimuth. The ionosphere is assumed to be two-dimensionally
inhomogeneous (over the propagation direction and over the height).

\subsection{Stage 2: Clustering by Dummy Teacher}

In order to further train the classification algorithm based on the
initially unlabeled data, within the framework of this algorithm,
it is necessary to create a labeled training dataset. Sometimes, the labeling
of the source dataset can be made manually, but it takes too much efforts
to do this. In this paper we use another approach, developed for
optimal embeddings in language models: search for the representation (hidden classes) 
optimal for solving a model problem by another network, and train
two these networks together. As a model problem, the imitation
of clustering by Dummy Teacher is used. Thus, the Dummy
Teacher is not the main element of the network, but rather a supporting one, which
allows, however, to optimally train Classifier network.

Scattered signals usually can be separated into a types according
to the boundary conditions of some measured parameter: these can
be short ranges for meteor trail scattering or low velocity and spectral
width for a groundscatter \cite{Nishitani_2019}. Therefore, it looks reasonable to use
for clustering the algorithms that finds an optimal limited areas occupied
for points of a given class. We used for clustering the probabilistic model,
describing the data as a superposition
of 20 multidimensional Gaussian distributions with their own parameters. Each of gaussian
describes its own cluster of points in multidimensional coordinates. This clustering
algorithm is widely known as Gaussian mixture model.

The parameters that the Dummy Teacher will use, in addition to
the physical parameters obtained as a result of simulation in stage
1, are the following experimental features: time (UT), spatial coordinates
(beam number or azimuth, radar distance, elevation of the received
signal) and the sounding frequency. For a more accurate
separation during clustering, we use only signals with a high signal-to-noise
ratio ($> 3dB$). The power of the radio signal is not used in the algorithm
due to the data contain various power variations
that are not related to the type of signal: radio signal absorption
in the D-layer and its focusing–defocusing due to the propagation
of the radio wave in an inhomogeneous ionosphere \cite{Berngardt_2021}.

Clustering of each experiment (the experiment number is a unique number
composed of the day number and the radar number) is carried out independently,
which allows us to speed up the clustering algorithm and put an additional
uncertainty to the Dummy Teacher.

\subsection{Stage 3: Training the Wrapped Classifier}

Our main task is to build and train the Classifier network, which
carries out the optimal classification of our data. The Classifier
architecture is shown in Fig.\ref{fig:1}B.

As the input of this network the physical parameters are used, which
we consider to be responsible for the signals identification, taking into
account the physical mechanisms of the formation of signals of different
types. This allows us to interpret the obtained classes in the future.

Typically, Doppler velocity and spectral width are used to identify
groundscatter signals. In addition to them, the scattering height,
which is responsible for the echo formation mechanism, carries useful
information. In particular, at altitudes below 100 km, the main signal source
is scattering by meteor trails; at altitudes up to 400 km
this is scattering in the ionosphere, and the higher altitudes
more likely corresponds to incorrect calculation of radiowave trajectory.
The Classifier
also uses parameters obtained during numerical simulation: key parameters
characterizing the propagation trajectory, the aspect angle with the Earth's magnetic
field at the scattering point, the real scattering height, and so
on.

It should be noted that the parameters used by the Dummy Teacher and
by the Classifier are different. The Classifier uses only physical
parameters that do not depend on the operating mode of the radar,
with which the optimal classification should be associated: scattering
height, trajectory angle with the magnetic field in the scattering
region, trajectory behavior features, propagation hop number, Doppler
frequency shift and signal spectral width. Dummy Teacher uses, in addition
to the physical parameters, also characteristics related
to the time and operating mode of the radar: the universal time, 
radar beam number, ray elevation, radar
range and sounding frequency.

Thus, the Classifier network make an extraction of meaning from the
input data and produces a vector (embedding) characterizing the classification
of signals from physical point of view, into hidden classes optimal to predict Dummy
Teacher clustering. In order to this embedding has an interpretable
probability meaning, it is normalized: the sum of all coordinates
of the vector is 1, and all the coordinates are non-negative.
This is implemented using the standard machine learning approach -
softmax activation function at the output of the Classifier.
This allows us to interpret this embedding as the probability
that the input point belongs to one or another "hidden" class. This embedding
will be used later by the "Wrap" network to produce prediction of
Dummy Teacher cluster number. The dimension of
this embedding vector is fixed and set to 20 in order to exceed
the possible number of significant types in the received signal.
As a result of training this dimension will be decreased automatically
(some of embedding coordinates will be nearly zero),
as it will be shown later.

The schema is trained based on the best fit by the Wrapped Classifier
the clusterization of the Dummy Teacher and algorithmically
formulated as classification training, where labels come from Dummy
Teacher and their predictions come from Wrapped Classifier.

The architecture of the network is shown
in Fig.\ref{fig:1}A.
Such an architecture allows training the Wrapped Classifier to classification
optimal from the point of view of predicting the results of a Dummy
Teacher. 
This architecture automatically renumerates the
cluster numbers for each experiment independently.
When processing new data, we do not need to use Wrap, but use only Classifier network and its
output (hidden classes probabilities) as physics-based classification of the signals. So
after training Wrapped Classifier the Wrap can be thrown away.

During training, weighted cross-entropy is used as a loss function,
where the weights are the inverse frequency of a cluster number in the training
dataset. This allows to automatically balance the dataset, and thereby to improve fitting quality.

AURPC is used as a metric of the prediction quality, because it works
correct enough in a case of a possible class imbalance. When training,
we use a early stopping over 20 epochs, when maximum AURPC is found
at validation dataset, as an overtrain stopping criterion. Gradient
descent method is used with ADAM optimizer, with batch size 32. Full dataset size
is about 3 million records, split into training, validation and test
datasets is made as 64\%:16\%:20\%. The propagation parameters are
calculated in the C language using the IRI-2012 and IGRF models, implemented
in the Fortran language. Network training is implemented in Python,
Dummy Teacher is implemented using sklearn library functions, Wrapped
Classifier is implemented using TensorFlow/Keras library functions.
The AURPC value achieved as a result of training was 0.659 after 173
training epochs, which looks as a sufficient result for a 20-class
classification.

\section{Results and their interpretation}

\subsection{Training results}

\begin{figure}
\includegraphics[scale=0.5]{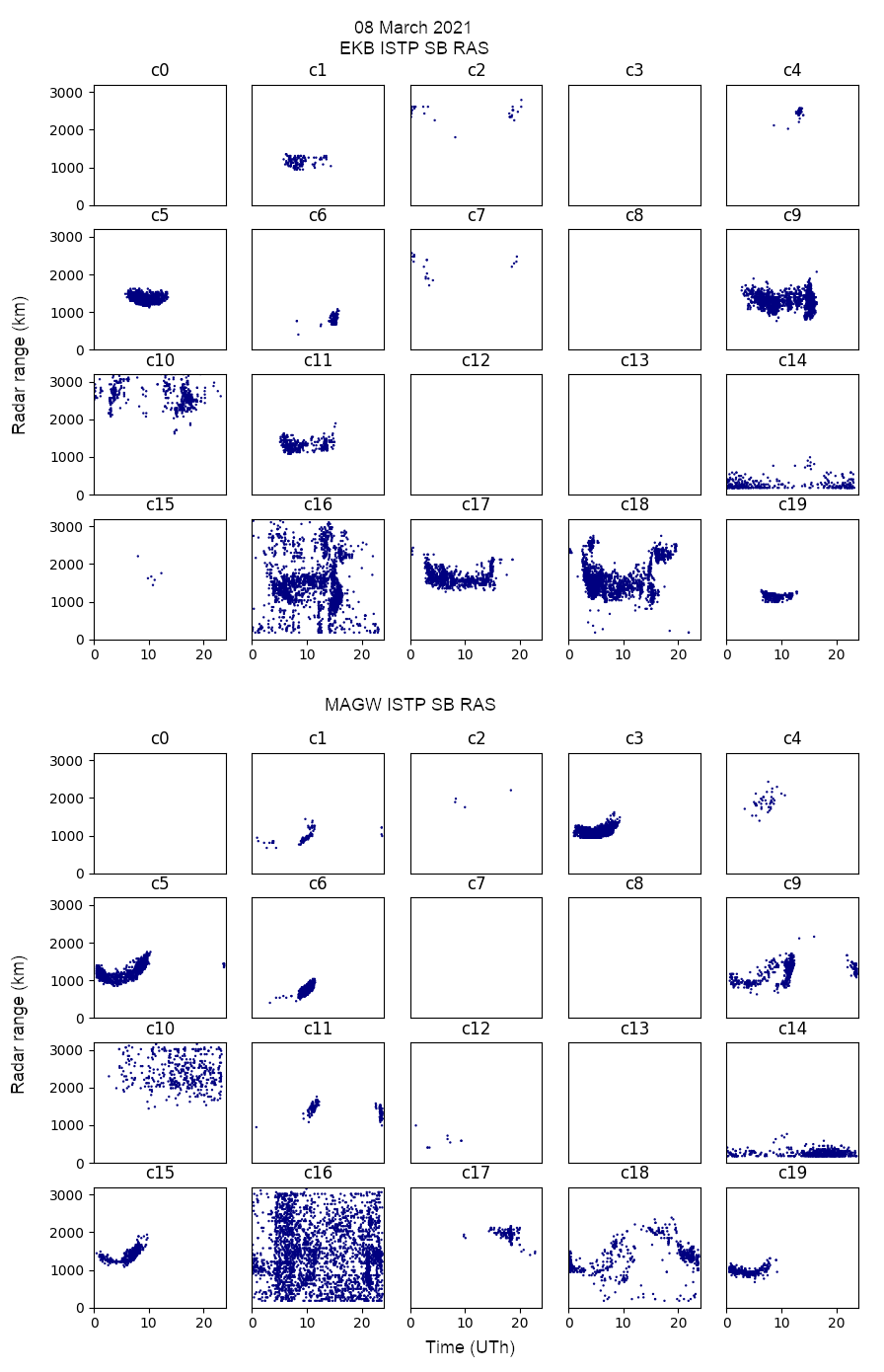}
\caption{Points positions of hidden classes of the Classifier as a function of radar range and time.
Data from EKB and MAGW ISTP SB RAS radars during 08/03/2021.}
\label{fig:all2}
\end{figure}

Fig.\ref{fig:all2} shows the result of classifying the 03/08/2021
data obtained by EKB and MAGW radars by the trained Classifier algorithm
into 20 classes (c0-c19). The data from all radar beams are displayed
simultaneously. The figure shows a reasonable separation of the data into classes.
Meteor scatter (class 14) is one of the most easily interpreted classes,
corresponding to scatter at radar ranges up to 450 km \cite{Berngardt_2020}.

\subsection{Interpreting the classes}

To interpret all the classes, the mean value and standard deviation of
each of the 10 Classifier input parameters and radar distance are
calculated for each class over whole available dataset (January-September 2021). 
According to the average values and standard
deviations of these parameters, the interpretation of each class is
carried out. In addition the degree of interpretability of
the classes is calculated - the number of points in which the mean
height of their classes does not exceed the physically expected value
of 400 km - the maximum of the F2 layer height. Higher values indicate
inaccurate ionospheric model or inaccurate received signal elevation angle measurements, 
which leads to errors in calculating
the wave propagation trajectory. The part of data that cannot be interpreted
from this point of view is 22\%, and there are 78\% of interpretable data points.

Classes 0,2,7,10,17,18 due to the high model scattering height exceeding
the height of the ionospheric maximum were assigned by us to the classes
with poorly calculated propagation trajectory. Scattering at high altitudes most
likely does not correspond to reality because of the low background
electron density at these altitudes, which leads to a low probability
of intense irregularities at altitudes above F2 maximum. Therefore, most likely,
this signal is scattered at lower altitudes, so the ray-tracing trajectory
calculations are most likely incorrect. This
is associated with known problems of the accuracy of ionospheric
models\cite{Liu_2019}, and with the inaccuracy of measuring the elevation angle
of the radio wave arrival\cite{Berngardt_2020}. 

Class 8 is empty, and classes 2, 4, and 13 are relatively rare, so
they should not be taken into account.

Classes 1,11 are localized at ranges of 800-1500 km, at heights of
35-200 km, non-aspect sensitive (the average angle with
the magnetic field is far from perpendicular), on the descending part
of the propagation trajectory. Therefore they can be interpreted as
long-range (1hop, after reflection from ionosphere) meteor scattering
or as a groundscatter.

Class 3 - rarely observed by the EKB radar, but frequently observed by the MAGW radar,
non-aspect sensitive, at heights of 50-110 km, at ranges of
1000-1250 km, after reflection from the ground on the ascending part
of the trajectory. It can be interpreted as 1.5hop meteor scatter.

Class 5 - non-aspect sensitive scattering at altitudes of 20-110 km,
after reflection from the ground, on the ascending part of the trajectory.
It can be interpreted as a groundscatter.

Class 6 - aspect sensitive scattering from distances of 500-800 km,
at heights of 160-260 km on the ascending part of the trajectory. It can
be interpreted as 0.5E or 0.5F scatter.

Class 9 - non-aspect sensitive scattering at ranges 600-1600 km, at
altitudes 40-180 km, on the descending part of the trajectory. It
can be interpreted as a groundscatter.

Class 12 - aspect sensitive scattering 
at altitudes of 200-300 km and ranges of 380-700 km, on the
ascending part of the trajectory. It can be interpreted as 0.5F scatter.

Class 14 is scattering at heights of 50-130 km, at distances of 180-470
km, on the ascending part of the trajectory. It can be interpreted
as a 0.5 hop meteor trail echo.

Class 15 - aspect sensitive scattering at heights of 125-190 km, 1100-1650
km, after reflection from the ground, on the ascending part of the
trajectory, can be interpreted as 1.5F scatter.

Class 16 - aspect sensitive scatter at heights of 0-550 km, at distances
500-2000 km, partially before, partially after reflection from the
ground. As one can see from Fig.\ref{fig:all2} it can be interpreted
as an indefinite class, consistent from points difficult to identify.
This class will not be taken into account.

Class 19 - non-aspect sensitive scattering at altitudes of 10-120 km,
on the descending part of the trajectory, at ranges of 1000-1200 km,
can be interpreted as groundscatter.

\begin{figure}
\includegraphics[scale=0.22]{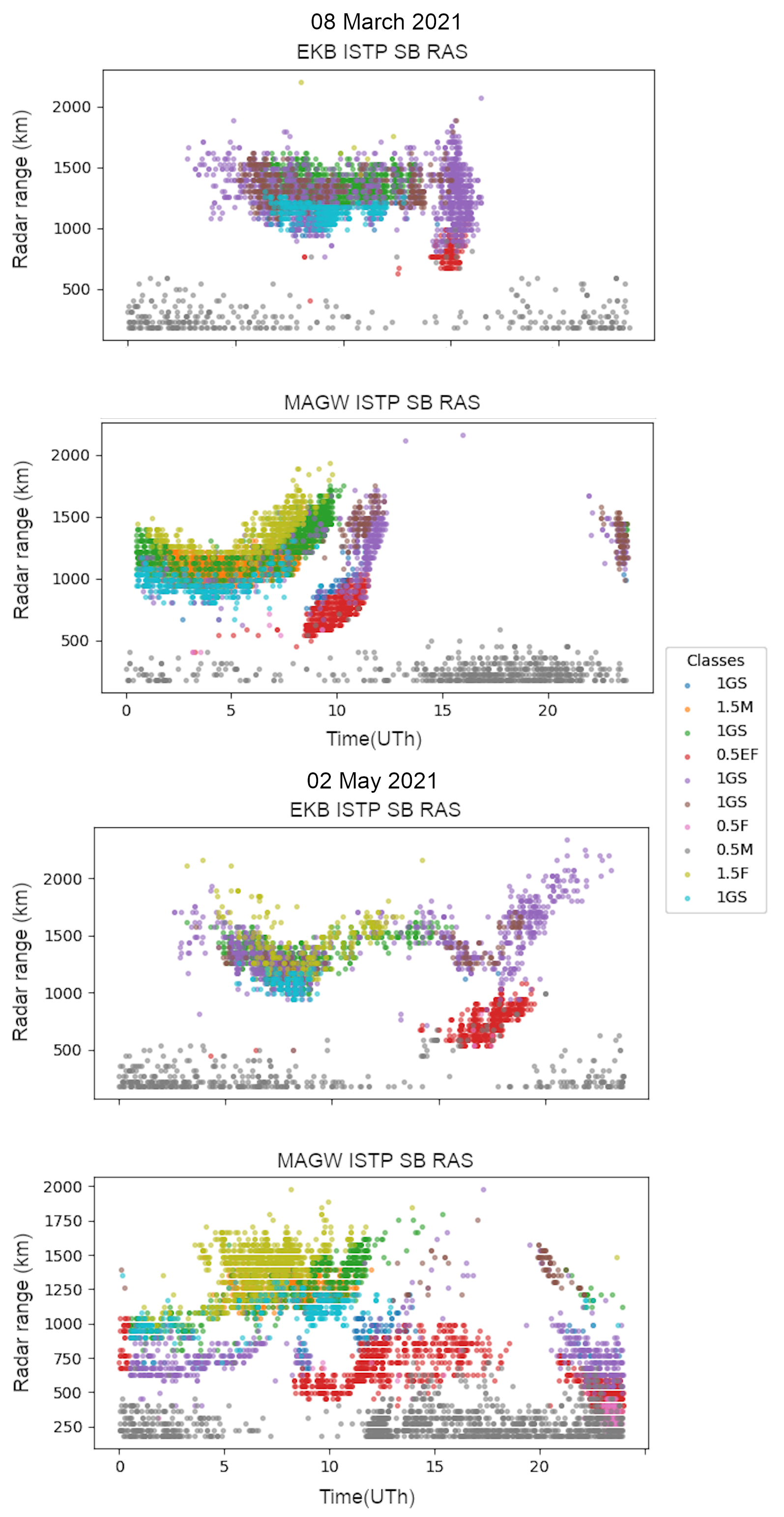}
\caption{An example of automatic classification of radar data EKB ISTP SB RAS
and MAGW ISTP SB RAS for 08 March 2021 (top) and 02
May 2021 (bottom)}
\label{fig:total2}
\end{figure}

Thus, the scattered signal can be divided into 11 interpretable classes, 
and 9 classes should not be taken into account.
An example of observations labeled in accordance
with this classification is shown in the Fig.\ref{fig:total2}. The
figure shows only the signals from interpreted classes.

\subsection{Simplifying the network}

To investigate the possibilities of simplifying the Wrapped Classifier
network, its simplified variants were analyzed: removing the straightening
space (PF) before Wrapped Classifier, removing one of the blocks (B2)
from the Classifier, removing the dependence on the refraction-free scattering height
($h_{eff}$). As an additional information,
the ability of the algorithm to separate meteor trail scattering into
a separate class on both radars was estimated (this is
determined from the average physical parameters of the class
- the signals of this class should have an average altitude of 70-100 km and an average radar
range not exceeding 400 km). The analysis results are summarized in
the table \ref{tab:1}.

\begin{table}
\caption{Comparison of results for simplified network variants. PF - Polynomial features
transformation is included into Wrap, B2 - Block 2 included into Classifier,
$h_{eff}$ - refraction free scattering height is used as input parameter.
Interpretable - percent of points in classes with correct scattering
height. ME - meteor trail echo.}

\begin{tabular}{|c|c|c|c|c|c|}
\hline 
PF & B2 & $h_{eff}$ & AURPC & Interpretable & ME class exists?\tabularnewline
\hline 
\hline 
yes & yes & yes & 0.659 & 78\% & yes\tabularnewline
\hline 
yes & yes & no & 0.593 & 77\% & no\tabularnewline
\hline 
yes & no & yes & 0.627 & 85\% & no\tabularnewline
\hline 
yes & no & no & 0.575 & 89\% & no\tabularnewline
\hline 
no & yes & yes & 0.550 & 77\% & no\tabularnewline
\hline 
no & yes & no & 0.522 & 74\% & no\tabularnewline
\hline 
no & no & yes & 0.605 & 86\% & no\tabularnewline
\hline 
no & no & no & 0.556 & 78\% & no\tabularnewline
\hline 
\end{tabular}\label{tab:1}
\end{table}

As can be seen from the results, the presented version of the Wrapped
Classifier network provides the best AURPC value, a good percentage
of class interpretability, and is the only variant that provides the
detection of meteor trail scattering by both radars. Thus, it looks
like no simplification of the network architecture is necessary.

\section{Conclusion}

In the paper a method for automatic classification of signals received
on EKB and MAGW ISTP SB RAS radars during January-September 2021 is described. The method is suitable
for automatic physical interpretation of the experimental data classification 
in realtime. We called this algorithm Wrapped
Classifier with Dummy Teacher. During training, it classifies
the radar data to 20 hidden classes, used for optimal prediction 
the data clustering by Dummy Teacher (the probabilistic
method of dividing data on 20 clusters with different gaussian distributions).

We extended parameters obtained by the radar with
additional parameters, calculated during numerical simulation of radiowave propagation
using ray-tracing technique and IRI-2012 and IGRF models for ionosphere and
Earth's magnetic field correspondingly.
For clustering by Dummy Teacher algorithm we use the whole dataset
of available parameters (measured and simulated ones). For classification
by Wrapped Classifier algorithm it uses only well physically interpreted
parameters.
As a result we found 11 well-intepretable classes from a physical point
of view in the available data.

Necessary for this method is a high-quality elevation calibration
of the radars and high-accuracy ionospheric model. 
They are necessary for correct calculations of the signal
propagation trajectory. For simulation we use the model ionosphere 
(IRI-2012) and ray-tracing
calculation of beam propagation in the ionosphere. Errors associated
with this lead to the appearance of classes that cannot be interpreted
from the point of view of radio wave propagation. They are characterized
by a large expected scattering height, above the maximum of the
F2 layer. The number of such uninterpretable datapoints is about
22\%.

\section*{Acknowledgments}

EKB ISTP SB RAS facility from Angara Center for Common Use of scientific
equipment (http://ckp-rf.ru/ckp/3056/), the radars are operated under
budgetary funding of Basic Research program II.12. The data of EKB
and MAGW ISTP SB RAS radars are available at ISTP SB RAS (http://sdrus.
iszf.irk.ru/ekb/page\_example/simple). 
The data analysis was performed in part on the equipment of the Bioinformatics Shared Access Center, the Federal Research Center Institute of Cytology and Genetics of Siberian Branch of the Russian Academy of Sciences (ICG SB RAS).
The authors are grateful to I.S.Petrushin (Irkutsk State University) for useful discussions.
The work has been done under
financial support of RFBR grant \#21-55-15012.

\bibliographystyle{IEEEtran}
\bibliography{IEEEabrv,biblio}


\end{document}